\documentclass[prl,showpacs,amsmath,amssymb,twocolumn,aps]{revtex4}

\usepackage{graphicx,graphics}
\usepackage{epsfig}
\usepackage{dcolumn}
\usepackage{bm}
\usepackage{natbib}
\usepackage{times}
\usepackage{ulem}
\def\kmm#1  {{\bf [KMM:~ #1]~}}
\def\new#1 {{\bf #1 }}
\def\cut#1 {\sout{#1} }

\newcommand{\pmn}{PMN~J0134$-$0931}

\newcommand{\dal}{\ensuremath{\Delta \alpha/ \alpha}}

\newcommand{\dmu}{\ensuremath{\Delta \mu/\mu}}

\newcommand{\beq}{\begin{equation}}
\newcommand{\eeq}{\end{equation}}

\newcommand{\lsb}{\left[}
\newcommand{\rsb}{\right]}

\begin{document}
\title{Constraints on changes in fundamental constants from a cosmologically
distant OH absorber/emitter}

\author{N.~Kanekar$^1$}
\email{nkanekar@aoc.nrao.edu}
\author{C.~L.~Carilli$^1$}
\author{G.~I.~Langston$^1$}
\author{G.~Rocha$^2$}
\author{F.~Combes$^3$}
\author{R.~Subrahmanyan$^4$}
\author{J.~T.~Stocke$^5$}
\author{K.~M.~Menten$^6$}
\author{F.~H.~Briggs$^{4,7}$}
\author{T.~Wiklind$^{8}$}
\affiliation{$^1$National Radio Astronomy Observatory, Socorro, NM 87801, USA}
\affiliation{$^2$Cavendish Laboratory, Cambridge CB3 0HE, UK}
\affiliation{$^3$Observatoire de Paris, F-75014, Paris, France}
\affiliation{$^4$Australia Telescope National Facility, Epping, NSW 1710, Australia}
\affiliation{$^5$University of Colorado, Boulder, CO 80309, USA}
\affiliation{$^6$Max-Planck-Institut f\"ur Radioastronomie, 53121 Bonn, Germany}
\affiliation{$^7$Australian National University, ACT 2611, Australia}
\affiliation{$^8$Space Telescope Science Institute, Baltimore, MD 21218, USA }

\date{\today}

\begin{abstract}
We have detected the four 18cm OH lines from the $z \sim 0.765$ gravitational lens 
toward \pmn. The 1612 and 1720~MHz lines are in conjugate absorption and 
emission, providing a laboratory to test the evolution of fundamental
constants over a large lookback time. We compare the HI and OH~main line absorption
redshifts of the different components in the $z \sim 0.765$ absorber and the
$z \sim 0.685$ lens toward B0218+357 to place stringent constraints on
changes in $F \equiv g_p \lsb \alpha^2/\mu \rsb^{1.57}$. We obtain
$\lsb \Delta F/F\rsb = (0.44 \pm 0.36^{\rm stat} \pm 1.0^{\rm syst}) \times 10^{-5}$,
consistent with no evolution over the redshift range $0 < z \lesssim 0.7$.  The 
measurements have a $2 \sigma$ sensitivity of $\lsb \dal \rsb < 6.7 \times 10^{-6}$ 
or $\lsb \dmu\rsb < 1.4 \times 10^{-5}$ to fractional changes in 
$\alpha$ and $\mu$ over a period of $\sim 6.5$~Gyr, half the 
age of the Universe. These are among the most sensitive constraints on changes in $\mu$.
 \end{abstract}

\maketitle
{\it Introduction.}--- A fairly generic feature of modern higher-dimensional 
theoretical models is that fundamental constants like the fine structure constant 
$\alpha$, the electron-proton mass ratio $\mu \equiv m_e/m_p$, the proton 
gyromagnetic ratio $g_p$, etc, depend on the scale lengths of the extra 
dimensions of the theory. In the current theoretical framework, it is implausible that 
these scale lengths remain constant, implying that $\alpha$, $\mu$, etc 
should vary with time. The detection of such changes provides an avenue to 
probe new and fundamental physics.

Null results have been obtained in all terrestrial studies of evolving
constants, with atomic clocks and isotopic abundances in the Oklo natural
fission reactor providing the tightest constraints on fractional
changes in the fine structure constant ($(1/\alpha)\lsb \Delta \alpha/{\Delta t} 
\rsb < 4 \times 10^{-15}$ per year, over three years \cite{peik04}, and
$\dal < 1.2 \times 10^{-7}$, over $\sim 1.8 \times 10^9$ years \cite{damour96},
respectively). However, terrestrial measurements only probe fairly small 
fractions of the age of the Universe; astrophysical techniques 
are needed to examine the possibility of variations at earlier times
(e.g. \cite{murphy03,srianand04,kanekar04,tzanavaris05,ivanchik05}).
It is these techniques that provide tantalizing evidence for changes in $\alpha$;  
the ``many-multiplet'' method, applied to Keck telescope optical 
spectra, gives $\dal = (-5.4 \pm 1.2) \times 10^{-6}$ over the redshift range 
$0.2 < z < 3.7$ \cite{murphy03}. However, a similar technique, applied to Very 
Large Telescope spectra, yields a conflicting result,
$\dal = (-0.6 \pm 0.6) \times 10^{-6}$, over $0.4 < z < 2.3$
\cite{srianand04}. Independent techniques are clearly needed as systematics
appear to play a significant role in the current results.

The four 18cm radio OH lines have very different dependences on $\alpha$, 
$\mu$ and $g_p$ and their redshifted frequencies can hence be compared to each 
other and to those of the HI 21cm hyperfine line or HCO$^+$ rotational lines 
to measure any variation \cite{chengalur03,darling03}. 
Even more interesting is the case of conjugate emission/absorption by the 18cm 
satellite OH lines, detected in a single cosmologically distant object (at $z \sim 0.247$ 
toward PKS1413+135; \cite{kanekar04,darling04}). Here, the pumping mechanism
guarantees that the two lines arise from the same gas; a comparison can thus be made
between the 1720 and 1612~MHz redshifts without concerns about systematic motions
between the clouds in which the different lines arise. 

Only four redshifted OH main absorbers are currently known \cite{kanekar02}, with 
high resolution data only available for the $z\sim 0.685$ lens toward 
B0218+357 \cite{kanekar03}.  Similarly, only one redshifted conjugate 
OH satellite system is known,
at a relatively low redshift, $z \sim 0.247$; this corresponds to a lookback
time of $\sim 2.9$~Gyr, not much earlier than the time range probed
by the Oklo reactor \footnote{Throughout this paper, we use an LCDM cosmology,
with $\Omega_m = 0.3$, $\Omega_\Lambda = 0.7$ and $H_0 = 70$~km/s~Mpc$^{-1}$.}.
We report here the detection of all four 18cm OH lines in a new conjugate
system at $z \sim 0.765$, corresponding to a lookback time of $\sim 6.7$~Gyr.

\begin{figure*}[t!]
\centering
\epsfig{file=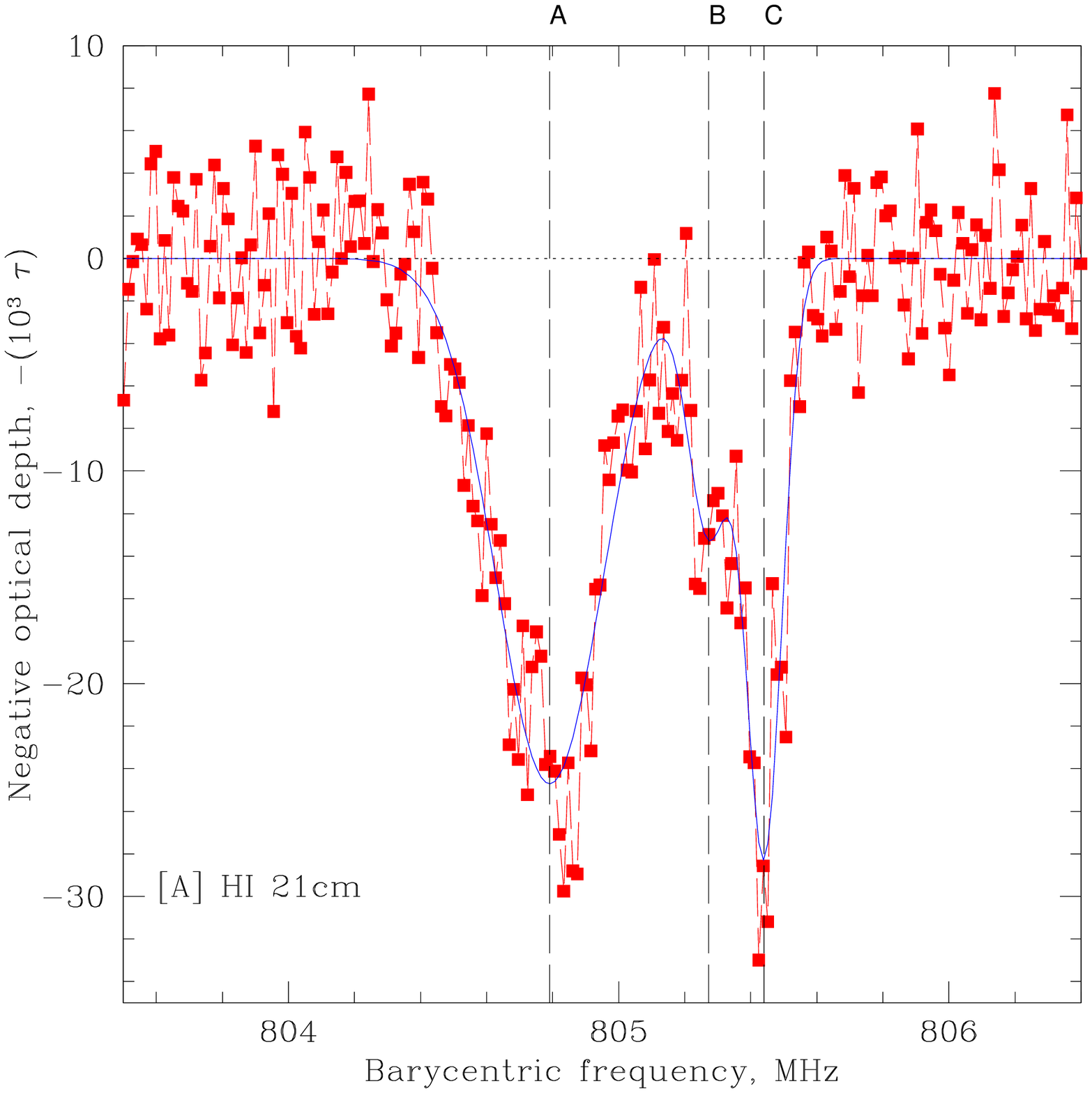,height=3.3in,width=3.3in}
\epsfig{file=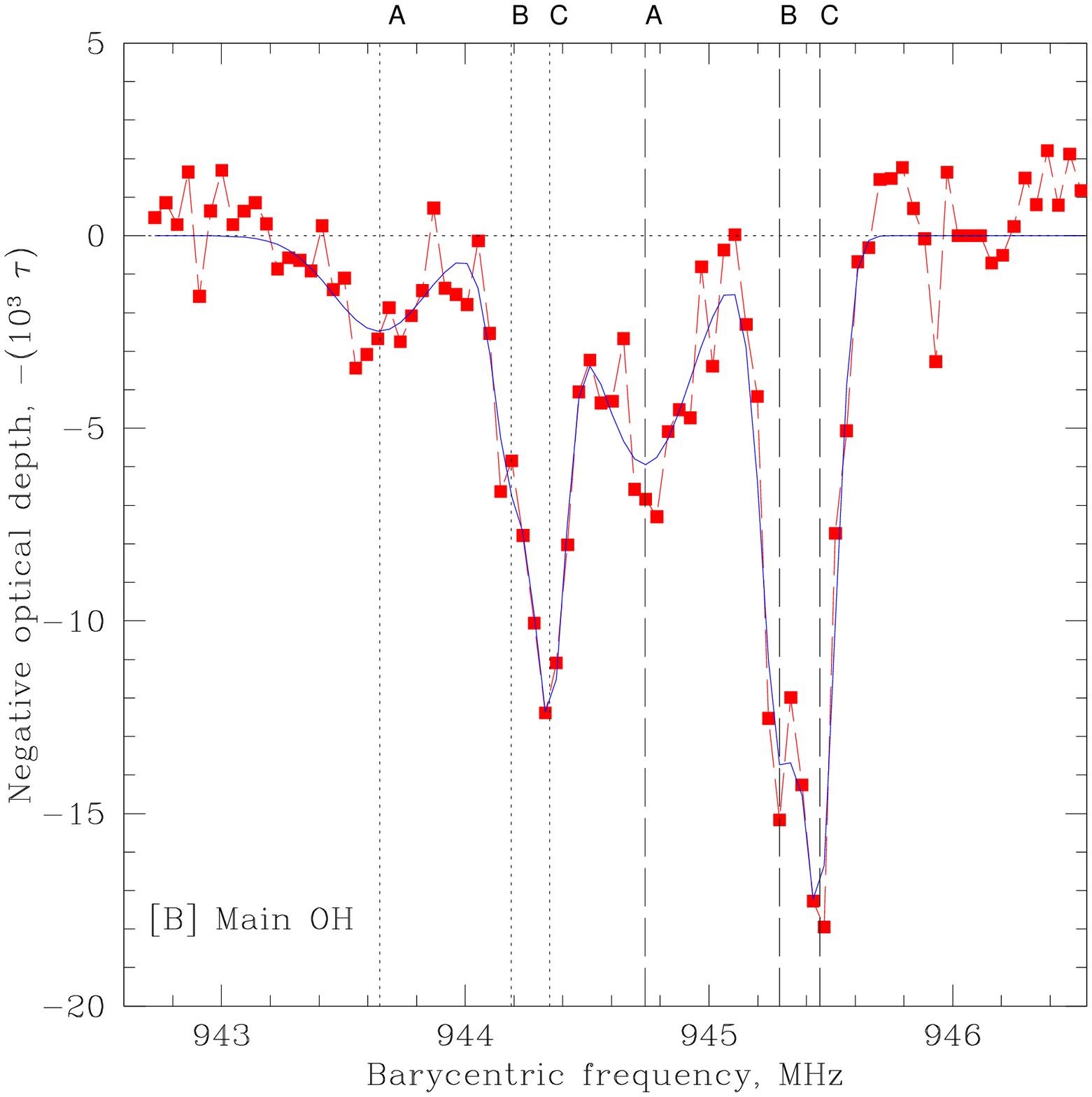,height=3.3in,width=3.3in}
\vskip -0.1in
\caption{GBT HI~21cm ([A], $\sim 5$~km/s resolution) and OH main line 
([B], $\sim 14.5$~km/s resolution) spectra
toward \pmn, with negative optical depth $(-10^3 \times \tau)$ plotted against
barycentric frequency, in MHz. The solid line shows the three-gaussian fit to each
spectrum. The vertical lines in each figure indicate the locations of the three
components (marked A, B and C), with the dashed and dotted lines in [B]
showing the 1667 and 1665 components, respectively. }
\vskip -0.2in
\label{fig:hi-oh}
\end{figure*}

\begin{figure}
\centering
\hskip -2.5 in \epsfig{file=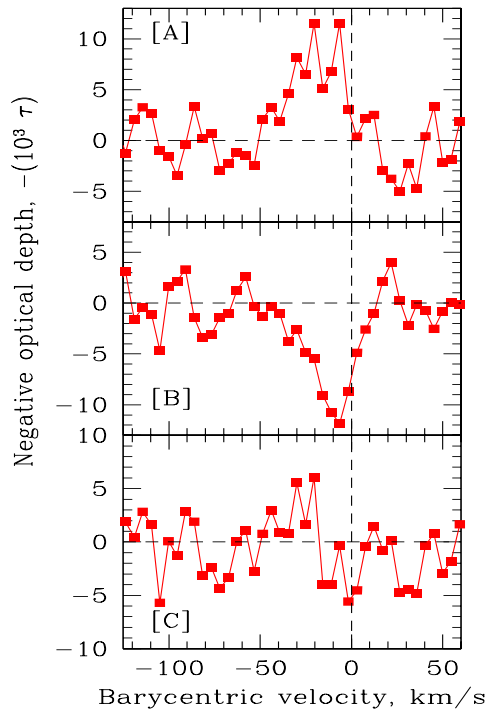,height=4.0in,width=5.0in}
\vskip -0.1in
\caption{OH satellite line spectra ($\sim 5$~km/s resolution), with negative optical depth ($-10^3 \times 
\tau $) plotted against barycentric velocity, in km/s, relative to $z = 0.76355$, 
the redshift of component C.
	[A] 1720~MHz transition, redshifted to $\sim 975.6$~MHz; [B] 1612~MHz transition,
	      redshifted to $\sim 912.2$~MHz;
          [C] Sum of 1612 and 1720~MHz spectra; this spectrum is
                consistent with noise. Note that the feature at $\sim -30$~km/s
		is not statistically significant. The satellite lines peak 
		$\sim 10$~km/s blueward of component C (the dashed vertical line), 
		which has the lowest redshift of the three spectral components.}
\vskip -0.2in
\label{fig:oh-sat}
\end{figure}

{\it Spectra and results}---
The redshifted OH 18cm lines from the $z \sim 0.765$ lens toward \pmn~
were observed simultaneously with the Green Bank Telescope (GBT) in October, 2004,
and January, 2005, with an observing resolution of $\sim 1$~km/s (after Hanning 
smoothing). A separate 
GBT observation in January, 2005, provided a high resolution spectrum in the 
redshifted HI~21cm line, originally detected by \cite{kanekar03b}. The 
HI~21cm and main line OH spectra (smoothed to resolutions of $\sim 5$ and 
$\sim 14.5$~km/s, respectively, and resampled) are shown in Figs.~\ref{fig:hi-oh}[A] 
and [B], while the three panels of Fig.~\ref{fig:oh-sat} show the 1720 and 
1612~MHz satellite spectra and the sum of 1720 and 1612~MHz optical depths 
(all at a resolution of $\sim 4.7$~km/s). 
\pmn~is unresolved by the GBT beam; the above optical depths are 
the ratio of line flux density to continuum flux density for 
each transition (using the low optical depth limit).
All spectra have a root-mean-square optical depth noise of $\sim 0.0018$, per 
$\sim 5$~km/s channel.

Besides the above, the redshifted HCO$^+$ 2--1 line was observed with the
IRAM 30m telescope and the Australia Telescope Compact Array (ATCA), the 6~cm ground state
H$_2$CO doublet lines with the GBT and the 2~cm first rotationally
excited state H$_2$CO lines
with the Very Large Array and the GBT. None of these transitions were detected,
down to  $3\sigma$ limits of $\tau < 0.07$ for the HCO$^+$ line and
 $\tau < 0.002$, $\tau < 0.005$ for the ground and excited state H$_2$CO lines,
respectively. This is
curious as every other redshifted OH absorber has also been detected in
HCO$^+$ absorption (e.g.~\cite{wiklind98,kanekar02}).
While the structure of the background source is very different at the HCO$^+$ 
and OH frequencies, implying that small-scale structure in the molecular cloud 
could be an issue, it is still surprising that none of the
OH components, at very different velocities from each other, show any trace of 
mm-wave absorption.

The HI~21cm profile has three fairly clear components, two of which are blended
and well-separated from the third. Similarly, both the 1667 and 1665~MHz OH lines
show two clearly resolved components, with the lower redshift one somewhat
asymmetric, suggesting that it is blended. While the redshifts of the two strong
21cm and main OH components are similar, the OH satellite lines are
blue-shifted by $\sim 10$~km/s relative to the lowest redshift (highest frequency)
component in Fig.~\ref{fig:hi-oh}[A] and [B]. This is reminiscent of the situation in the
other redshifted conjugate OH system, PKS1413+135 \cite{kanekar04}.  In both cases,
the sum of the satellite frequencies is different from the sum of the main
line frequencies; since the two sums have the same dependence on $\alpha$, $\mu$
and $g_p$ \cite{chengalur03,darling03}, the satellite and main OH lines must
arise in different gas.

The sum of the 1720 and 1612~MHz optical depths in Fig.~\ref{fig:oh-sat}[c] 
is consistent with noise; the satellite lines are thus conjugate
with each other.  Such conjugate behavior arises due to competition between the
intra-ladder $119\mu$ and cross-ladder $79\mu$ decay routes to the OH ground state,
after the molecules have been pumped by collisions or far-infra-red radiation into
the higher excited states (e.g. \cite{langevelde95}). The fact that the 1720~MHz
line is seen in emission and the
1612~MHz, in absorption, implies that the intra-ladder decay route dominates. This,
and the requirement that the $119\mu$ transition be optically thick
for the 1720~MHz line inversion \cite{langevelde95}, yields the constraint
$3.5 \times 10^{15} \lesssim N_{\rm OH} \lesssim 3.5 \times 10^{16}$~cm$^{-2}$
on the OH column density.

The 1720~MHz line luminosity is $L_{\rm OH} \sim 3700 L_\odot$, making this
the brightest known 1720~MHz megamaser by more than an order of magnitude. It is
also the most distant OH megamaser; for comparison, the furthest known 1667~MHz 
megamaser is at $z \sim 0.265$ \cite{baan92}. This is also the first case 
of conjugate OH satellite emission/absorption in a ``normal''
galaxy; all previous cases were objects containing an active galactic nucleus
(e.g. Cen~A; \cite{langevelde95}), where the OH level populations might have been 
affected by nuclear activity.  This is interesting as it suggests that such conjugate 
behaviour might not be as rare as earlier expected and hence, that it might be used 
as a tool to probe both spatial and temporal changes in fundamental constants.

{\it Constraining changes in fundamental constants.}---
A comparison between the redshifts of different spectral lines to measure
changes in fundamental constants involves the assumption that the lines have
no intrinsic velocity offsets from each other. This is true even for comparisons
between lines of the same species as different transitions might be excited
under different physical conditions and thus, in different spatial locations.
This appears to be the case with the main and satellite OH lines here, implying
that one cannot compare their redshifts to estimate changes in the
different constants.

However, as in the case of PKS1413+135 \cite{kanekar04,darling04}, the conjugate 
nature of the satellite lines ensures that they arise in the same
physical region and, crucially, that systematic velocity offsets are not an
issue. The different dependences of the sum and difference of the 1720~MHz and 1612~MHz
frequencies on $\alpha$, $\mu$ and $g_p$ then allows us to estimate
changes in the quantity $G \equiv g_p \lsb \alpha^2/\mu \rsb^{1.85}$
\cite{chengalur03,kanekar04}.  The low signal-to-noise ratio of the 1720
and 1612 spectra of Fig.~\ref{fig:oh-sat} precludes such an estimate at the
present time. However, the high redshift of the system implies that
it is an excellent target for deep integrations in the satellite lines, enabling
a precision measurement of changes in $\alpha$, $\mu$ and $g_p$ in the 
future. For example, observations with the Square Kilometer Array,
a next generation radio telescope, should be able to detect
fractional changes $\Delta G / G \sim 5 \times 10^{-7}$ in this system, implying
a sensitivity of $\dal \sim 1.4 \times 10^{-7}$. This is similar to
the sensitivity of the Oklo measurement but with fewer assumptions and out to
a far larger lookback time of $\sim 6.7$~Gyr. In addition, this sensitivity 
would be obtained from a single system, unlike the many-multiplet method, which 
requires a large number of absorbers to average out systematic effects. 
A comparison between the results from the conjugate 
systems in PKS1413+135 and \pmn~ will thus allow one to probe true spatio-temporal 
changes in the above constants (instead of merely averaging over spatial effects), 
especially since the two sources are very widely separated on the sky.

Comparisons between the HI 21cm and main OH lines suffer from the drawback of
possible systematic velocity offsets between the two species.
However, a tight correlation has been found between the velocities of HCO$^+$ and nearest
HI~21cm absorption in the Galaxy, with a dispersion of only $\sim 1.2$~km/s
\cite{drinkwater98}. Further, the velocities of Galactic OH and HCO$^+$ absorption
have been found to be remarkably similar (\cite{liszt00}; see Figs.~4 
and 5 of \cite{liszt00}). OH and HI~21cm velocities should thus also be well-correlated 
and, in fact, the dispersion in this correlation is likely to be less than that between 
the HCO$^+$ and HI~21 velocities, as the spatial structure of the background source is
quite similar at the nearby OH~18~cm and HI~21cm frequencies. It should thus be possible
to use a comparison between main OH and HI~21cm absorption from a statistically 
large number of redshifted systems as an independent probe of any evolution 
in $\alpha$, $\mu$ and $g_p$ \cite{chengalur03}. We next apply this technique 
to the absorbers 
toward \pmn~ and B0218+357, albeit using a more conservative dispersion of 3~km/s 
between OH and HI intrinsic velocities, characterising internal motions within a 
molecular cloud.

The HI~21cm profile of Fig.~\ref{fig:hi-oh}[A] has the highest signal-to-noise
ratio of our spectra (at the same resampled resolution of $\sim 1$~km/s) 
and has three clear components. We hence used a three-gaussian template to locate 
the peak redshifts of the different 21cm absorption components, with the amplitudes, 
positions and widths of the gaussians all left as free parameters. A similar 
three-gaussian fit was then carried out to the smoothed and resampled 1665 and 
1667~MHz spectra, with the difference that the velocity widths here were
fixed to those obtained from the 21cm fit and only the amplitudes and
positions of individual components left as free parameters. The 1665 
and 1667~MHz fits were carried out simultaneously, to account for the
possibility of blending between the components. Note that the 
original spectral resolution was $\sim 1$~km/s in all cases and no additional 
components were seen in any of the spectra. The possibility that strong narrow 
components might be blended in the lower resolution spectra of Figs.~\ref{fig:hi-oh}~[A] 
and [B] can thus be ruled out.

Figs.~\ref{fig:hi-oh}[A] and [B] show the fitted gaussians as solid lines,
overlaid on the HI and OH spectra. The three fitted 21cm components have peak redshifts 
$z_{\rm HI-A} = 0.764938 \pm 0.000015$, $z_{\rm HI-B} = 0.763881 \pm 0.000033$ and 
$z_{\rm HI-C} = 0.763515 \pm 0.000014$, while the sums of the 1665 and 1667~MHz 
frequencies have peak redshifts $z_{\rm OH-A} = 0.764871 \pm 0.000042$,
$z_{\rm OH-B} = 0.763852 \pm 0.000019$ and $z_{\rm OH-C} = 0.763550 \pm 0.000010$.
Comparing these redshifts, component by component, we obtain
$\lsb \Delta F/F\rsb_{\rm A} = (-3.7 \pm 2.8) \times 10^{-5}$,
$\lsb \Delta F/F\rsb_{\rm B} = (-1.6 \pm 2.1) \times 10^{-5}$ and
$\lsb \Delta F/F\rsb_{\rm C} = (2.0 \pm 1.0) \pm 10^{-5}$, where
$F \equiv g_p \lsb \alpha^2/\mu \rsb^{1.57}$ and only statistical errors 
have been included. A weighted average of these values gives $\lsb \Delta F/F \rsb =
(0.86 \pm 0.86) \times 10^{-5}$. Note that a comparison using an entirely unconstrained 
six-Gaussian fit to the OH spectrum gives the weighted average $\lsb \Delta F/F \rsb = 
(2.25 \pm 0.84) \times 10^{-5}$. Similarly, the HI~21cm and main OH redshifts 
in the $z \sim 0.685$ lens toward B0218+357 are $z_{\rm HI} = 0.684676 \pm 0.000005$ 
\cite{carilli00} and $z_{\rm OH} = 0.684682 \pm 0.0000056$ \cite{chengalur03}, 
giving $\lsb \Delta F/F \rsb_{\rm D} = 
(3.5 \pm 4.0) \times 10^{-6}$. All the above values are consistent with 
the null hypothesis of no evolution in the different constants. Combining 
results from the two absorbers (using the constrained fit and a weighted average), 
we obtain $\lsb \Delta F/F\rsb = (0.44 \pm 0.36^{\rm stat} \pm 1^{\rm syst}) \times 10^{-5}$, 
over $ 0 < z \lesssim 0.7$, where the second error is due to velocity offsets 
between HI and OH lines, assuming a velocity dispersion of 3~km/s. 
Of course, four measurements are far too few for a reliable estimate of 
this error. However, the fact that the OH and HI redshifts are in reasonable 
agreement within the measurement errors in all four cases (two of which have 
errors $\lesssim 1 \times 10^{-5}$) suggests that systematic velocity offsets do 
not dominate the accuracy of the measurement. 

The strong dependence of $F$ on $\alpha$ and $\mu$ ($F \propto \alpha^{3.14}$ 
and $F \propto \mu^{-1.57}$) implies a $2 \sigma$ sensitivity of
$\lsb \dal \rsb < 6.7 \times 10^{-6}$ or $\lsb \dmu\rsb < 1.4 \times 10^{-5}$
to fractional changes in $\alpha$ and $\mu$ from $z \sim 0.7$ (i.e. a
lookback time of $\sim 6.5$~Gyrs) to today, where we have added the errors in
quadrature (these sensitivities are not independent as we measure $\lsb \Delta F/F \rsb$). 
Assuming linear evolution, these correspond to $2\sigma$
sensitivities of $(1/\alpha) \lsb \Delta \alpha/ \Delta t \rsb < 1.1 \times
10^{-15}$~yr$^{-1}$ or $(1/\mu) \lsb \Delta \mu/\Delta t \rsb < 2.1 \times
10^{-15}$~yr$^{-1}$, among the best present sensitivities to changes in 
$\mu$. For comparison, \cite{ivanchik05} obtains 
$\lsb \dmu \rsb < 1.48 \times 10^{-5}$ for $0 < z \lesssim 2.75$ while
\cite{tzanavaris05} obtains $\lsb \dmu\rsb \lesssim 2 \times 10^{-5}$ 
for $ 0 < z < 2.0$, at $2\sigma$ level, using redshifted optical lines.
The present radio analysis is not affected by two important sources 
of systematic error in the optical regime (wavelength 
calibration and relative isotopic abundances; e.g.  \cite{murphy03}),
although, of course, the optical observations probe a larger redshift range. 
The primary source of the error in our technique is likely to lie in the 
velocity dispersion 
between OH and HI lines and perhaps in blending between weak narrow 
spectral components. We do not feel that these dominate the present 
results; deeper observations in the OH and HI lines should help 
quantify their effects.

While the size of the radio sample is still small, surveys are being carried 
out that will significantly increase the number of known redshifted OH, HI and 
HCO$^+$ absorbers. Comparisons between radio lines are thus likely to provide 
an important independent constraint on changes in fundamental constants in the 
future.

\begin{acknowledgments}
We thank Bob Garwood and Jim Braatz for help with the data analysis.
The National Radio Astronomy Observatory is operated by Associated Universities, 
Inc., under cooperative agreement with the National Science Foundation. The
Australia Telescope Compact Array is part of the Australia Telescope, funded by 
the Commonwealth of Australia for operation as a National Facility managed by CSIRO.

\end{acknowledgments}

\bibliography{ms4}

\end{document}